\documentstyle[12pt,titlepage,epsbox]{article}
\begin{document}
\pagestyle{empty}
\baselineskip=0.212in

\begin{flushleft}
\large
{SAGA-HE-78-95
\hfill January 12, 1995}  \\
\end{flushleft}

\vspace{1.5cm}

\begin{center}

\Large{{\bf  Nuclear Shadowing }} \\

\vspace{0.3cm}

\Large{{\bf  in the Structure Function $\bf F_3(x)$ }} \\

\vspace{1.2cm}

\Large
{R. Kobayashi, S. Kumano, and M. Miyama $^*$ }         \\

\vspace{0.6cm}

\Large
{Department of Physics, Saga University, Saga 840, Japan } \\

\vspace{1.5cm}

\Large{ABSTRACT}

\end{center}

Nuclear modification of the structure function $F_3$ is investigated.
Although it could be estimated in the medium and large $x$ regions
from the nuclear structure function $F_2^A$,
it is essentially unknown at small $x$.
The nuclear structure function $F_3^A$ at small $x$ is investigated
in two different theoretical models:
a parton-recombination model with $Q^2$ rescaling
and an aligned-jet model.
We find that these models predict completely different
behavior at small $x$: {\it antishadowing} in the first parton model
and {\it shadowing} in the aligned-jet model.
Therefore, studies of the ratio $F_3^A/F_3^D$ at small $x$
could be useful in discriminating among different models,
which produce similar shadowing behavior in the structure function $F_2$.
We also estimate currently acceptable nuclear modification of
$F_3$ at small $x$ by using $F_2^A/F_2^D$ experimental data and
baryon-number conservation.

\vspace{0.6cm}

\vfill

\noindent
{\rule{6.cm}{0.1mm}} \\

\vspace{-0.2cm}
\normalsize
\noindent
{* Email: kobar, kumanos, or 94sm10@cc.saga-u.ac.jp.
   Information on our research is}  \\

\vspace{-0.6cm}
\noindent
{available at
http://www.cc.saga-u.ac.jp/saga-u/riko/physics/quantum1/structure.html} \\

\vspace{-0.6cm}
\noindent
\normalsize
{or at ftp://ftp.cc.saga-u.ac.jp/pub/paper/riko/quantum1.} \\

\vspace{1.0cm}

\vspace{-0.2cm}
\hfill
{submitted for publication}

\vfill\eject
\pagestyle{plain}

\noindent
{\Large\bf {1. Introduction}}
\vspace{0.4cm}

It is known that nuclear structure functions are not
identical to the corresponding ones in the nucleon.
The nuclear modification was first found by
the European Muon Collaboration (EMC), so that it is called
EMC effect.
The phenomena have been well investigated theoretically and
experimentally in the structure function $F_2$ \cite{EMCF2,SKF2}.
In recent years, the small $x$ region has been
studied extensively.
The ratio $R_2\equiv F_2^A/F_2^N$ \cite{COMM} becomes smaller than
unity at small $x$, which is referred to as
shadowing in the structure function $F_2$.
Models for explaining the $F_2$ shadowing
include vector-meson-dominance-type models
and parton-recombination-type models \cite{SKREF}.

The former models describe the shadowing in the following way.
A virtual photon transforms into vector-meson states
(or $q\bar q$ states), which then interact
with a target nucleus.
The central constituents are ``shadowed" due to the existence
of nuclear surface constituents.
The latter models explain the shadowing by interactions
of partons from different nucleons in a nucleus,
and the interactions are called parton recombinations.
They become important especially at small $x$, where
the longitudinal localization size of a parton exceeds
the average nucleon separation in the nucleus.

Because these two different ideas produce similar shadowing results
in $F_2$, it is difficult to distinguish
among the models in comparison with experimental data.
If accurate experimental $F_2$ data are obtained \cite{NMCNEW},
it is perhaps possible to find the ``correct" explanation.
However, a better idea could be to look for other
shadowing observables.
We discuss in this paper that
a promising way of testing the models
could be to use the structure function $F_3$.

Except for the nuclear effects on $F_2$,
modification of sea-quark and
gluon distributions \cite{SKGLUE} has been discussed
recently.
However, there is another interesting point to be investigated.
It is the structure function $F_3$ in nuclei.
In a naive parton model without next-to-leading-order corrections,
$F_3$ is given by valence-quark distributions.
Therefore, the ratio $R_3\equiv F_3^A/F_3^N$
in the medium $x$ region should be well estimated
from existing experimental data for $R_2$.
So the essential point of our investigation is to study the small
$x$ region. Although there is a certain restriction
on $R_3$ due to baryon-number conservation,
the $F_3$ shadowing is an undeveloped research area.

The purpose of our research is to investigate
whether or not studies of the $F_3$ shadowing
help us discriminate among different models,
which produce similar results on the $F_2$ shadowing.
In section 2, nuclear shadowing in $F_3$ is investigated
in two models:
1) a parton recombination model with $Q^2$ rescaling effects
and
2) an aligned-jet model.
Then, using the existing data of $R_2$ and the conservation rule,
we estimate an acceptable $F_3$ shadowing region.
Our conclusions are given in section 3.

\vfill\eject
\noindent
{\Large\bf {2. Shadowing in the structure function $\bf F_3(x)$}}
\vspace{0.4cm}

We discuss model predictions of
the nuclear-structure-function ratio $R_3$.
In particular, two different models are employed.
The first model is the parton-recombination model with
$Q^2$ rescaling in Ref. \cite{SKF2}.
The second model is the aligned-jet model by
Frankfurt, Liuti, and Strikman  \cite{AJM,AJMF3}.
This model is based on the vector-meson-dominance model
but it is modified due to $q\bar q$ continuum.

\vspace{1.0cm}
\noindent
{\bf 2.1 A parton-recombination model with $\bf Q^2$ rescaling}
\vspace{0.4cm}

After the EMC finding of nuclear modification in 1983,
there are many publications on the topic.
There is no unique explanation for the EMC effect unfortunately.
However, it is the fact that similar results are obtained
in completely different models, for example,
in a macroscopic model of nuclear binding
and in a microscopic model of $Q^2$ rescaling.
This duality is interpreted in the following way \cite{RGR,RESCALE}.
As there is freedom to choose the renormalization point
in an operator product expansion,
the factorization scale in convolution formalism is not defined.
A different factorization scale corresponds
to a different interpretation of the EMC effect.
In this sense, we could connect seemingly different models
by a single scale change, which is called rescaling.
According to Ref. \cite{RESCALE}, nuclear structure functions
$F_2^A(x,Q^2)$ are given by rescaling $Q^2$
in the nucleon structure function $F_2^N(x,Q^2)$.

We use this $Q^2$ rescaling model with parton recombinations \cite{SKF2}.
The rescaling model was originally proposed as a model
for explaining the medium $x$ region, and the recombination \cite{RECOM}
as a model in the small $x$ region.
Using these two mechanisms, we explained
the structure function $F_2^A$ from very small $x$ to large $x$
\cite{SKF2}.
This unified model can be applied to the structure function $F_3$.
In the naive parton model,
the $F_3$ structure function of the nucleon
is given by $F_3(x)=xu_v(x)+xd_v(x)$.
The formalism in Ref. \cite{SKF2} is used in calculating
the valence-quark distributions.
The initial $Q^2$ was chosen at 0.8 GeV$^2$ in calculating
the rescaling and the recombinations; however, the resulting
evolution shows little $Q^2$ dependence in the ratio $R_2$.
Calculated results at $Q^2$=0.8 GeV$^2$
are shown in Fig. 1 for the calcium nucleus.
The dashed curve shows the $Q^2$ rescaling results.
The ratio $R_3$ in the medium $x$ region is almost equal
to the EMC effect in $F_2$.
As $x$ increases, the ratio becomes smaller and typical
nuclear effects are 10\% in the medium size nucleus.
Because the $Q^2$ rescaling satisfies the baryon-number conservation,
the ratio $R_3$ becomes larger as $x$ decreases.
There are other important effects at small $x$ due to
parton recombinations.
Their contributions on $F_3(x)$ are rather contrary to
those in the $Q^2$ rescaling model as shown in Fig. 1.
The recombinations decrease the ratio at small $x$ and
increase it at medium and large $x$.
In the model of Ref. \cite{SKF2}, the valence-quark
number is conserved, so that the ratio is above unity
at small $x$ due to the medium $x$ suppression.
The overall nuclear modification is very interesting
in the sense that the ratio $R_3$ is much different from
the one for the structure function $F_2(x)$ at small $x$.
In this parton model, the $F_3$ shadowing differs
distinctively from the $F_2$ one:
$$
{{F_3^A(x)} \over {F_3^N(x)}} \ne
{{F_2^A(x)} \over {F_2^N(x)}} ~~~~~{\rm at}~~{\rm small}~~x  ~~~.
\eqno{(2.1)}
$$
In other words, valence-quark modification is different from
the sea-quark one. It is especially interesting to find in Fig. 1
that the model predicts {\it antishadowing} in
the structure function $F_3$ instead of shadowing.

\vspace{1.0cm}
\noindent
{\bf 2.2 An aligned-jet model}
\vspace{0.4cm}

Within the parton model in subsection 2.1,
the $F_3$ modification at small $x$ is very different from the $F_2$ one.
However, the situation is not so simple.
There exists a model in which $F_2$ and $F_3$ shadowing
predications are similar.
It is an aligned-jet model, for example, the one investigated
in Ref. \cite{AJM,AJMF3}.
It is based on a traditional idea, the vector-meson-dominance model.
The virtual photon transforms into vector-meson states, which
interact with a target.
The propagation length of the hadronic fluctuation is
estimated by $\lambda\approx 1/|E_H -E_\gamma| \approx 0.2/x$ fm.
The length exceeds the average nucleon separation in a nucleus
at small $x$, so that shadowing phenomena occur
due to multiple scatterings.

The aligned-jet model is an extension of this model.
The virtual photon (or W) transforms into a $q\bar q$ pair,
which then interacts with the target.
However, the only $q\bar q$ pair aligned in the direction
of $\gamma$ (W) interacts in a similar way
to the vector-meson interactions with the target.
In this model, vector-meson-like $q\bar q$ pairs
interact with sea quarks and valence quarks in the same manner.
Therefore, the $F_3$ shadowing is very similar to the $F_2$ one.
In fact, the shadowing results in figure 3
of  Ref. \cite{AJMF3} are similar to those of the $F_2$ shadowing:
$$
{{F_3^A(x)} \over {F_3^N(x)}} \approx
{{F_2^A(x)} \over {F_2^N(x)}} ~~~~~{\rm at}~~{\rm small}~~x  ~~~.
\eqno{(2.2)}
$$

It is interesting to find that the aligned-jet model predicts
the shadowing, which is contrary to the antishadowing
in the first parton model.
In the following, we discuss whether or not both possibilities
are allowed within existing data and the baryon-number conservation.

\vfill\eject
\vspace{1.0cm}
\noindent
{\bf 2.3 Experimental restriction and comparison}
\vspace{0.4cm}

We find that the calculated $F_3$ results
are much different in both models.
In this subsection, an experimental restriction
on the $F_3$ shadowing is discussed.
If next-to-leading-order effects are neglected,
the $F_3$ structure function is identical to
the valence-quark distribution:
$F_3(x)=u_v(x)+d_v(x) \equiv V(x)$.
We assume $R_V \equiv V_A(x)/V_N(x) =R_3$ in our investigation.
The $F_2$ structure function at medium and large $x$
is dominated by the valence-quark distributions.
Therefore, the ratio $R_3$ for the structure function $F_3$
at medium and large $x$ should be equal to $R_2$,
which has been measured experimentally.
The major point of our study is then to estimate
the small $x$ region.
There is a restriction on the valence-quark modification due to
the baryon-number conservation $\displaystyle{ \int dx V(x) =3}$.
Because the ratio $R_3$ is less than unity at medium $x$,
there are two major possibilities at small $x$.
One is that the ratio increases as $x$ decreases (antishadowing),
and the other is that the ratio is significantly enhanced
at $x\approx 0.1$ and it becomes smaller than unity
at very small $x$ (shadowing).

In order to study these possibilities, we assume that
the ratio $R_V$ at $x>0.3$ is equal to $R_2$
by neglecting sea-quark contributions.
SLAC experimental data $R_2$ \cite{SLACR2} of the calcium nucleus are
shown in Fig. 2 in the $x$ region, $x>0.3$.
These data are fitted by a smooth analytical function shown
by a solid curve.
We extrapolate the curve into the small $x$ region by considering
the baryon-number conservation.
First, a straight line in the logarithmic $x$ is simply drawn
from $x$=0.3 as shown by the dashed line in Fig. 2
so that it satisfies the conservation rule.
The line is roughly the upper limit of nuclear modification.
Second, the curve is smoothly extrapolated into the small $x$
region by allowing about 6\% antishadowing at $x=0.1-0.2$.
Considering experimental antishadowing in $F_2$ at $x=0.1-0.2$
\cite{NMC} and a nuclear sea-quark distribution \cite{SEA},
we think that 6\% is roughly an upper bound for the valence-quark
antishadowing in this region.
If accurate data are obtained in $R_2$ and
in the sea-quark-distribution ratio $S_A(x)/S_N(x)$ at $x=0.1-0.2$,
we could have a better estimate of the antishadowing.
Because of this antishadowing, the ratio could become smaller
than unity at very small $x$ in order to satisfy the conservation.
This dotted curve at small $x$
is considered to be a rough lower bound for the nuclear modification.
The shaded area between these curves is the area of possible
nuclear modification, which is allowed by present experimental
data of $R_2$ and the baryon-number conservation.
It should be emphasized that the region is a very naive
estimate of the experimental restriction.

It is noteworthy that the first parton-model (model 1)
prediction is roughly equal to the upper bound curve,
and the aligned-jet model (model 2)
prediction is to the lower bound curve.
So the models are two extreme cases, which are both
acceptable in our present knowledge.
We have not investigated the details of other model predictions.
However, it is very encouraging to investigate
the (anti)shadowing phenomena of $F_3$ in the sense that
the observable could be useful in discriminating among
different models, which produce similar results in
the $F_2$ shadowing.

There are reasonably accurate data for $F_3$ of a medium size nucleus,
but we do not have accurate deuteron data \cite{F3D}.
We hope that future neutrino experiments provide us
accurate information on nuclear modification of $F_3$.

\vspace{1.0cm}
\noindent
{\Large\bf {3. Conclusions}}
\vspace{0.4cm}

We investigated nuclear modification of the structure function $F_3$
in two different models: the parton-recombination with $Q^2$ rescaling
and the aligned-jet model.
These two models predict completely different behavior at small $x$:
antishadowing in the first model and shadowing in the second model.
Within our present experimental and theoretical knowledge,
both results are allowed even though they are two extreme cases.
Therefore, it is possible to rule out one possibility by measuring
$F_3^A/F_3^D$.
Furthermore, the $F_3$ shadowing could be useful in discriminating
among various shadowing models.
Our investigation is merely a starting point
in studying details of the $F_3$ shadowing.
In particular, other model predictions and experimental possibility
should be explored.

$~~~$

$~~~$

\noindent
{\Large\bf {Acknowledgments}}
\vspace{0.4cm}

S. K. would like to thank R. G. Roberts for discussion,
which motivated us to investigate the $F_3$ shadowing.
We thank A. Bodek for communications on the $F_3$ structure function.
This research was partly supported by the Grant-in-Aid for
Scientific Research from the Japanese Ministry of Education,
Science, and Culture under the contract number 06640406.

\vfill\eject


\vfill\eject
\noindent
{\Large\bf{Figure Captions}} \\

\vspace{-0.38cm}
\begin{description}
   \item[Fig. 1]
Nuclear modification of $F_3$ in the calcium nucleus
is predicted in the parton recombination model with $Q^2$ rescaling.
The dashed curve shows the results in the $Q^2$ rescaling model.
Effects of the parton recombinations are shown by the arrows.
The overall results including the rescaling and
the recombinations are shown by the solid curve.
   \item[Fig. 2]
The model-1 curve is the predication
of the nuclear modification of $F_3$ in the calcium
by the parton model in subsection 2.1,
and the model-2 curve is the one by the aligned-jet model.
The dashed and dotted curves are estimated by using
existing data of $F_2^{Ca}/F_2^D$ at $x>0.3$
and the baryon-number conservation.
The shaded area is roughly the currently acceptable region.
Note that the experimental data are for $F_2^{Ca}/F_2^D$.
\end{description}

\vfill\eject
\hspace{-3.5cm}
\epsfile{file=fig1.eps}

\vfill\eject
\hspace{-3.5cm}
\epsfile{file=fig2.eps}


\begin{thebibliography}{}
\bibitem{COMM} The nuclear structure function per nucleon is used
               throughout this paper.
\end{thebibliography}
\end{document}